\documentclass[twocolumn,epfs,epsfig,aps,nofootinbib,10pt]{revtex4-1}

\usepackage{graphicx}
\usepackage{pslatex}
\usepackage{epstopdf}
\usepackage{dcolumn}
\usepackage{amsmath}
\usepackage{float}

\begin{document}

\title{Numerical evolution of general Robinson-Trautman spacetimes: code tests, wave forms and the efficiency of the gravitational wave extraction}

\author{H. P. de Oliveira}
\email{hp.deoliveira@pq.cnpq.br}
\author{E. L. Rodrigues}
\email{elrodrigues@uerj.br}
\author{J. E. F. Skea}
\email{jimsk@dft.if.uerj.br}
\affiliation{{\it Universidade do Estado do Rio de Janeiro}\\
{\it Instituto de F\'{\i}sica - Departamento de F\'{\i}sica Te\'orica}\\
{\it CEP 20550-013 Rio de Janeiro, RJ, Brazil.}}

\date{\today}

\begin{abstract}
We present an efficient numerical code based on spectral methods to integrate the field equations of general Robinson-Trautmann spacetimes. The most natural basis functions for the spectral expansion of the metric functions are spherical harmonics. Using the values of appropriate combinations of the metric functions at the collocation points, we have managed to reduce expression swell when the number of spherical harmonics increases. Our numerical code  runs with relatively little computational resources and the code tests have shown excellent accuracy and convergence. The code has been applied to situations of physical interest in the context of Robsinson-Trautmann geometries such as: perturbation of the exterior gravitational field of a spheroid of matter; perturbation of an initially boosted black hole;  and the non-frontal collision of two Schwarzschild black holes. In dealing with these processes we have derived analytical lower and upper bounds on the velocity of the resulting black hole and the efficiency of the gravitational wave extraction, respectively. Numerical experiments were performed to determine the forms of the gravitational waves and the efficiency in each situation of interest.

\end{abstract}

\maketitle

\section{Introduction}

The use of numerical techniques to solve Einstein's field equations is a very promising strategy for dealing with problems of astrophysical interest such as gravitational collapse and formation of black holes, coalescence of binary systems, the collision of black holes, etc. One could say that numerical relativity is the bridge between relativity and astrophysics. A common feature in all these processes is the emission of gravitational waves - one of the most notable predictions of General Relativity.  Undoubtedly, detailed knowledge of how much system mass is converted into gravitational waves and the forms of the resulting gravitational waves are of fundamental importance in the efforts to detect directly gravitational waves with ground and space based observatories~\cite{gw_sources}.

In this context we have two main objectives. The first is to present an efficient numerical code for solving the field equations of general Robinson-Trautmann spacetimes,  based on spectral methods~\cite{spectral}. The second is to exhibit the initial results for the wave forms and efficiency of the gravitational wave extraction associated with the following systems: the nonlinear perturbation of an oblate distribution of matter evolving to form a black hole, the recoil of a boosted black hole after interacting with a packet of gravitational waves, and the non-frontal collision of two Schwarzschild black holes.

The Robinson-Trautman metric can be expressed as~\cite{rt}

\begin{eqnarray}
\label{eq1} ds^2&=&\left(\lambda - \frac{2m_0}{r} - 2 r
\frac{\dot{P}}{P}\right) d u^2 + 2 du dr \nonumber
\\
& & - \frac{r^2}{P^2}(d \theta^{2}+\sin^{2}\theta d
\phi^{2}),
\end{eqnarray}

\noindent where $u$ is a null coordinate such that $u={\rm constant}$ denotes null hypersurfaces generated by the rays of the gravitational field, and that foliates the spacetime globally; $r$ is an affine parameter defined along the null geodesics determined by the vector $\partial/\partial r$ and $m_0$ is a constant. The angular coordinates $(\theta,\phi)$ span the spacelike surfaces $u=$ constant, $r=$ constant commonly known as the 2-surfaces. In the above expression an overdot indicates a derivative with respect to $u$, the functions $\lambda$, identified as the Gaussian curvature of the 2-surfaces, and $P$ depend on the coordinates $u,\theta,\phi$. In the particular case of axial symmetry the metric functions do not depend on $\phi$. Two important aspects of Robinson-Trautman spacetimes are worth mentioning: they are asymptotically flat and admit the presence of gravitational waves.

Einstein's equations for the Robinson-Trautman (RT) spacetimes reduce to

\begin{eqnarray}
\label{eq2}
& & \lambda=P^2+\frac{P^2}{\sin \theta}\left(\sin \theta \frac{P_{,\theta}}{P}\right)_{,\theta}+\frac{P^2}{\sin^2 \theta}\left(\frac{P_{,\phi}}{P}\right)_{,\phi} \\
\nonumber \\
& & 12 m_{0}\frac{\dot{P}}{P}+P^2\left(\frac{(\lambda_{,\theta} \sin
\theta)_{,\theta}}{\sin \theta}+\frac{\lambda_{,\phi\phi}}{\sin^2\theta}\right)=0,\label{eq3}
\end{eqnarray}

\noindent where the subscripts $\theta,\phi$ denote derivatives with respect to the angles $\theta$ and $\phi$, respectively. The structure of the field equations is typical of the characteristic evolution scheme: Eq.~(\ref{eq2}) is the hypersurface equation that defines $\lambda(u,\theta,\phi)$, and Eq.~(\ref{eq3}), known as the RT equation, governs the dynamics of the gravitational field. In other words, from the initial data $P(u_0,\theta,\phi)$ the hypersurface equation determines $\lambda(u_0,\theta,\phi)$, and the RT equation allows us to evolve the initial data. 

Several works have been devoted to the evolution of the vacuum RT spacetimes focusing on the issue of the existence of solutions to the full nonlinear equation. The most general analysis on the existence and asymptotic behavior of the vacuum RT equation was given by Chrusciel and Singleton~\cite{chru}, and independently by Frittelli and Moreschi~\cite{moreschi1}. They proved that, for sufficiently smooth initial data, the spacetime exists globally for positive retarded times, and converges asymptotically to the Schwarzschild metric.

The combination of gravitational waves and the asymptotic convergence to the Schwarzschild solution allows RT spacetimes to be used to study the emission of gravitational waves in connection with the formation of a single black hole. The first work in this direction was due to Foster and Newman \cite{fn} who interpreted the linearized solution of the field equations as representing the emission of gravitational radiation by a bounded source. RT spacetimes have also been used as test beds for numerical codes in the characteristic formulation of axisymmetric spacetimes \cite{winicour}. The full evolution of the field equations (\ref{eq2}) and (\ref{eq3}) constitutes a valid and valuable framework for a detailed analysis of the wave forms and the efficiency of the gravitational wave extraction connected with the formation of black holes. Furthermore, a scenario in which the resulting black hole is moving with respect to an inertial observer due to, for example, the collision of two black holes, can be treated using the dynamics of RT spacetimes. Some of these aspects were addressed in Refs~\cite{oliv1,oliv2,oliv3,oliv4,oliv_rad,rod_saa,rod_rezzola} in the context of axial symmetry after the development of a numerical code for integrating the field equations~\cite{oliv_code}.

The paper is organized as follows. In Section~2 we present the details of our numerical scheme. In Section~3 we introduce the initial data employed for the code tests. Tests confirm the excellent accuracy and convergence of the numerical code. Section~4 is devoted to applying the code to situations of physical interest, with the gravitational wave forms and the efficiency of the gravitational wave extraction being evaluated after evolving the spacetime. In addition, by analyzing the conservation of the four-momentum, we were able to determine analytically a lower limit of the velocity of the black hole formed as well as an upper bound for the efficiency of the gravitational wave extraction. As a final and brief application, we consider initial data describing the collision of two Schwarzchild black  holes in three situations: orthogonal, oblique and head-on collisions. Finally, in Section 5 we present a summary of our results and conclusions.

\section{The numerical method}

According to the Galerkin method with numerical integration (G-NI)~\cite{galerkin}, the metric functions $P(u,\theta,\phi)$ and $\lambda(u,\theta,\phi)$ are approximated by a series expansions given by,

\begin{eqnarray}
P_N(u,\theta,\phi) = \sum_{k=0}^{N}\,\sum_{l=-k}^{l=k}\,a_{kl}(u) Y_{kl}(\theta,\phi) \label{eq4}\\
\nonumber \\
\lambda_N(u,\theta,\phi) = \sum_{k=0}^{N}\,\sum_{l=-k}^{l=k}\,b_{kl}(u) Y_{kl}(\theta,\phi), \label{eq5}
\end{eqnarray}

\noindent where $Y_{kj}(\theta,\phi)$ are the spherical harmonics chosen as the basis functions, $a_{kl}(u)$ and $b_{kl}(u)$ are the unknown modes and $N$ is the truncation order that dictates the number of independent modes. These modes must satisfy the symmetry relations $a^*_{k-l}=(-1)^{-l}a_{kl}$ and $b^*_{k-l}=(-1)^{-l}b_{kl}$ that result in a total number of $2 (N+1)^2$ independent modes. Although the number of independent modes increases considerably with the truncation order, the use of spherical harmonics provides exponential convergence for smooth functions on the sphere, and also fixes the so called pole problem~\cite{boyd}.

The substitution of the approximations Eqs.~(\ref{eq4}) and~(\ref{eq5}) into the field equations~(\ref{eq2}) and~(\ref{eq3}) results in the residuals associated to these equations,
{\small
\begin{eqnarray}
\mathrm{Res}_\lambda(u,\theta,\phi) &&= \lambda_M - P_N\,\sum_{k,l}\,(k^2+k-1)a_{kl} Y_{kl} + \nonumber \\
&& (P_{N,\theta})^2 + \frac{(P_{N,\phi})^2}{\sin^2 \theta}
 \label{eq6} \\
\nonumber \\
\mathrm{Res}_P(u,\theta,\phi)&&= 12\,m_0\,\sum_{k,l}\,\dot{a}_{kl} Y_{kl} - P_N^3 \times \nonumber \\
&&\sum_{k,l}\,k(k+1)b_{kl} Y_{kl}.
\label{eq7}
\end{eqnarray}
}

\noindent The modes ($a_{kj},b_{kj}$) are determined such that the residual vanish in the mean~\cite{finlayson}. This means that the inner products (projections) of the residuals with respect to the test functions are made equal to zero. For the G-NI method the test and basis functions are the same, then

\begin{eqnarray}
&&\left<\mathrm{Res}(u,\theta,\phi),Y_{mn}(\theta,\phi)\right> = \nonumber \\
&&\int_0^{2\pi}\,\int_0^\pi\,\mathrm{Res}(u,\theta,\phi)\,Y^{*}_{mn}\sin \theta d\theta d\phi = 0, \label{eq8}
\end{eqnarray}

\noindent where $\mathrm{Res(u,\theta,\phi)}$ stands for $\mathrm{Res}_\lambda(u,\theta,\phi)$ or $\mathrm{Res}_P(u,\theta,\phi)$, with $m$ varying from 0 to $N$, and $n=-m,..,m$.

As dictated by the G-NI method, these integrals were evaluated using quadrature formulas which provide fast and accurate results, since these formulas usually converge exponentially~\cite{boyd,quadr_form}. We introduce the variable $x \equiv \cos \theta $ and define the quadrature collocation points $(x_k,\phi_j)$ by

\begin{eqnarray}
\begin{array}{l l}
\phi_j = \frac{2 \pi}{\widehat{M}} \;\; j=0,1,..,\widehat{M} \\
\\
x_k =-1,\,\mathrm{zeros\,of}\, \mathcal{P}^\prime_{\widehat{N}}(x), 1,\label{eq9}
\end{array}
\end{eqnarray}

\noindent where $\mathcal{P}_{\widehat{N}}(x)$ is the Legendre polynomial of order~$\widehat{N}$, and we have set $\widehat{M}=2N$ and $\widehat{N}=3N/2$ that leads to a more accurate integration. The integration with respect to $\phi$ the trapezoidal rule was used that is the indicated quadrature rule for periodic functions \cite{boyd}, while the Legendre-Gauss-Lobatto quadrature formula was used for the integration with respect to $x$. These integrals can be  written schematically as

\begin{equation}
\int_0^{2\pi}\,\int_{-1}^1\,F(x,\phi)\, dx d\phi \approx \sum_{j=0}^{\widehat{M}}\sum_{k=0}^{\widehat{N}}\,v_jw_kF(x_k,\phi_j) \label{eq10}
\end{equation}

\noindent where $w_k,v_j$ are the weights~\cite{quadr_form}. 

In order to allow the use of high truncation orders, the projections of the residual equations~(\ref{eq6}) and~(\ref{eq7}) are expressed as,

\begin{eqnarray}
&& b_{kj} = S_{kj}\left(a_{lm},P_{lm},(P_{,\theta})_{lm},\left(\frac{P_{,\phi}}{\sin \theta}\right)_{lm}\right), \label{eq11}\\
\nonumber \\
&& \dot{a}_{kj} = F_{kj}(P_{lm},\bar{\lambda}_{lm}), \label{eq12}
\end{eqnarray}

\noindent where $P=P_N$ and the indices $lm$ indicate that the expression is evaluated at the collocation points $x=x_l,\phi=\phi_m$. These values are related to the modes $a_{kj}$ but stored in separated files. Also, $\bar{\lambda}_{lm}$ represent the values of $\bar{\lambda}(u,\theta,\phi) = \sum_{k,l}\,k(k+1)b_{kl}(u) Y_{kl}(\theta,\phi)$ at the collocation points. Therefore, the field equations (\ref{eq2}) and (\ref{eq3}) are reduced, respectively, to a set of algebraic equations connecting the modes $b_{jk},a_{lm}$, and a set of ordinary differential equations for the modes $a_{lm}$. The evolution scheme proceeds as follows: (i) from the initial data $P_0(x,\phi)=P(u_0,x,\phi)$ the initial modes $a_{jk}(u_0)$ are fixed from Eq. (\ref{eq4}); (ii) the values of $P,P_{,\theta}$ and $P_{,\phi}/\sin \theta$ at the collocation points are evaluated; (iii) the initial modes $b_{jk}(u_0)$ are determined from the set of equations (\ref{eq11}); (iv) the values of $\bar{\lambda}$ at the collocation points are evaluated; (v) the ordinary differential equations (\ref{eq12}) fix $\dot{a}_{jk}$ at $u=u_0$, and the modes $a_{jk}$ can be determined at the next null surface $u_0+\delta u$. By repeating the whole process the evolution of the modes is determined, and consequently the evolution of the functions $P(u,x,\phi)$ and $\lambda(u,x,\phi)$.

\section{Code tests}

The integration of the dynamical system (\ref{eq12}) requires $(N+1)^2$ initial conditions $a_{kj}(u_0)$, determined by the initial data $P_0(x,\phi)=P(u_0,x,\phi)$. We have determined physically relevant initial data for the axisymmetric case representing the exterior gravitational field of homogeneous and non-homogeneous spheroids~\cite{oliv_rad}, the perturbed boosted black hole by a gravitational wave packet~\cite{oliv4} and also the collision of two Schwarzschild black holes~\cite{oliv3}. We consider here the generalization of these initial data to the non-axisymmetric case, but details of the derivation will be omitted since they are straightforward.

We start with simple initial data which have no direct physical motivation,

\begin{eqnarray}
P_0^{(I)}(x,\phi)=\frac{\sqrt{1-e^2(1-x^2)\sin^2\phi}}{\sqrt{1-e^2}},
\label{eq13}
\end{eqnarray}

\noindent where $0<e<1$ plays the role of the eccentricity of the above spheroid-type function. The second set of initial data represents the exterior gravitational field of a perturbed oblate spheroid~\cite{oliv_rad},

\begin{equation}
P_0^{(II)}(x,\phi)=\left[1+\frac{B_0}{2}\left(\alpha+\frac{\beta}{4}(3x^2-1)\right) +  f(x,\phi)\right]^{-2},
\label{eq14}
\end{equation}

\noindent where the first two terms on the right describe a homogeneous oblate spheroid in which $B_0$ is an arbitrary parameter, $\alpha=\arctan(1/\zeta)$, $\beta=(1+3\zeta^2)\arctan(1/\zeta)$, with $\zeta$ being a free parameter associated with the oblateness of the spheroid~\cite{oliv_rad}. The function $f(x,\phi)$ can be interpreted as describing a perturbation inside the matter distribution of the spheroid. The third set of initial data is expressed as

\begin{eqnarray}
P_0^{(III)}(x,\phi) = \left(\frac{1}{\sqrt{P_0(\cosh \gamma \pm x \sinh \gamma)}}+g(x,\phi)\right)^{-2},
\label{eq15}
\end{eqnarray}

\noindent where $P_0$ and $\gamma$ are arbitrary parameters, and $g(x,\phi)$ is a regular function. This function generalizes the initial data of Ref.~\cite{oliv4} that describes a boosted black hole with respect to an asymptotic observer which is perturbed by a non-spherical mass distribution. The boosted black hole in the $z-$direction is characterized by the exact stationary solution~\cite{bondi} of the field equations (\ref{eq2}) and (\ref{eq3}),

\begin{equation}
P=P_0(\cosh \gamma \pm x \sinh \gamma),
\label{eq16}
\end{equation}

\noindent in which $\lambda=P_0^2$. The black hole is moving with velocity $v=\tanh \gamma$ with respect to an asymptotic observer~\cite{bondi}. This solution has been used to study the deceleration of a moving black hole due to an external perturbation and also the head-on collision of two black holes with the same velocity~\cite{oliv3}.

Due to the lack of exact non-stationary solutions of the field equations, the best way of testing the accuracy and convergence of the code is to verify if the conserved quantity~\cite{rt_austr}

\begin{equation}
I =\frac{1}{4 \pi} \int_0^{2\pi}\,\int_{-1}^1\,P^{-2}dx d\phi, \label{eq17}
\end{equation}

\noindent is maintained constant by the numerical solution within an acceptable error, and if the error diminishes when the truncation order $N$ increases. Since the initial data $P_0(x,\phi)$ determine the exact value, $I=I_{\mathrm{exact}}$, a useful measure of the numerical error with respect to the exact value of $I$ is

\begin{equation}
\delta I = \frac{|I_{\mathrm{exact}}-I_{\mathrm{numer}}|}{I_{\mathrm{exact}}},
\label{eq18}
\end{equation}

\noindent where $I_{\mathrm{numer}}$ is the numerical value of $I$.

\begin{figure}[htb]
\rotatebox{0}{\includegraphics*[height=4.5cm,width=5.2cm]{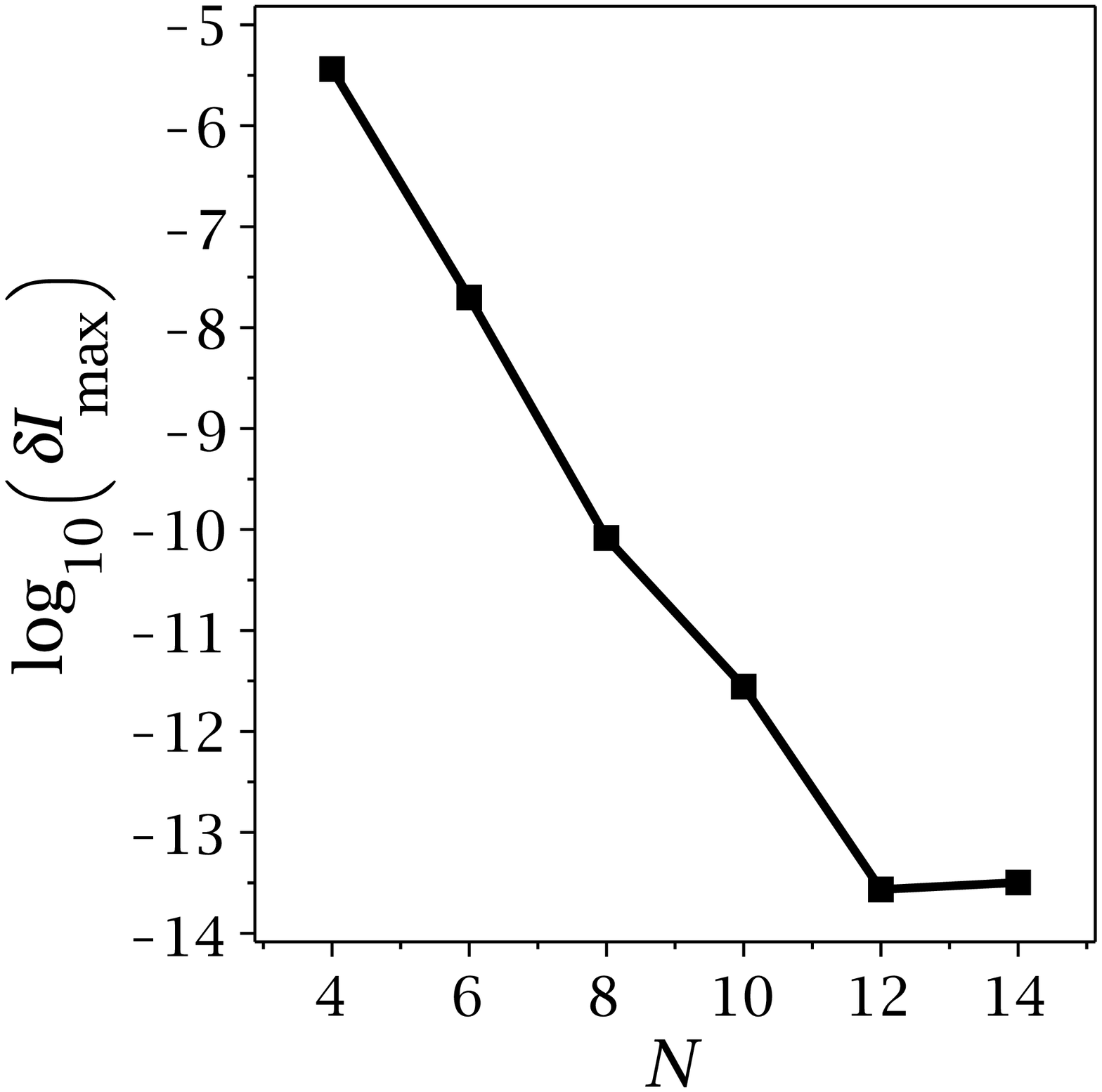}}
\rotatebox{0}{\includegraphics*[height=4.5cm,width=5.2cm]{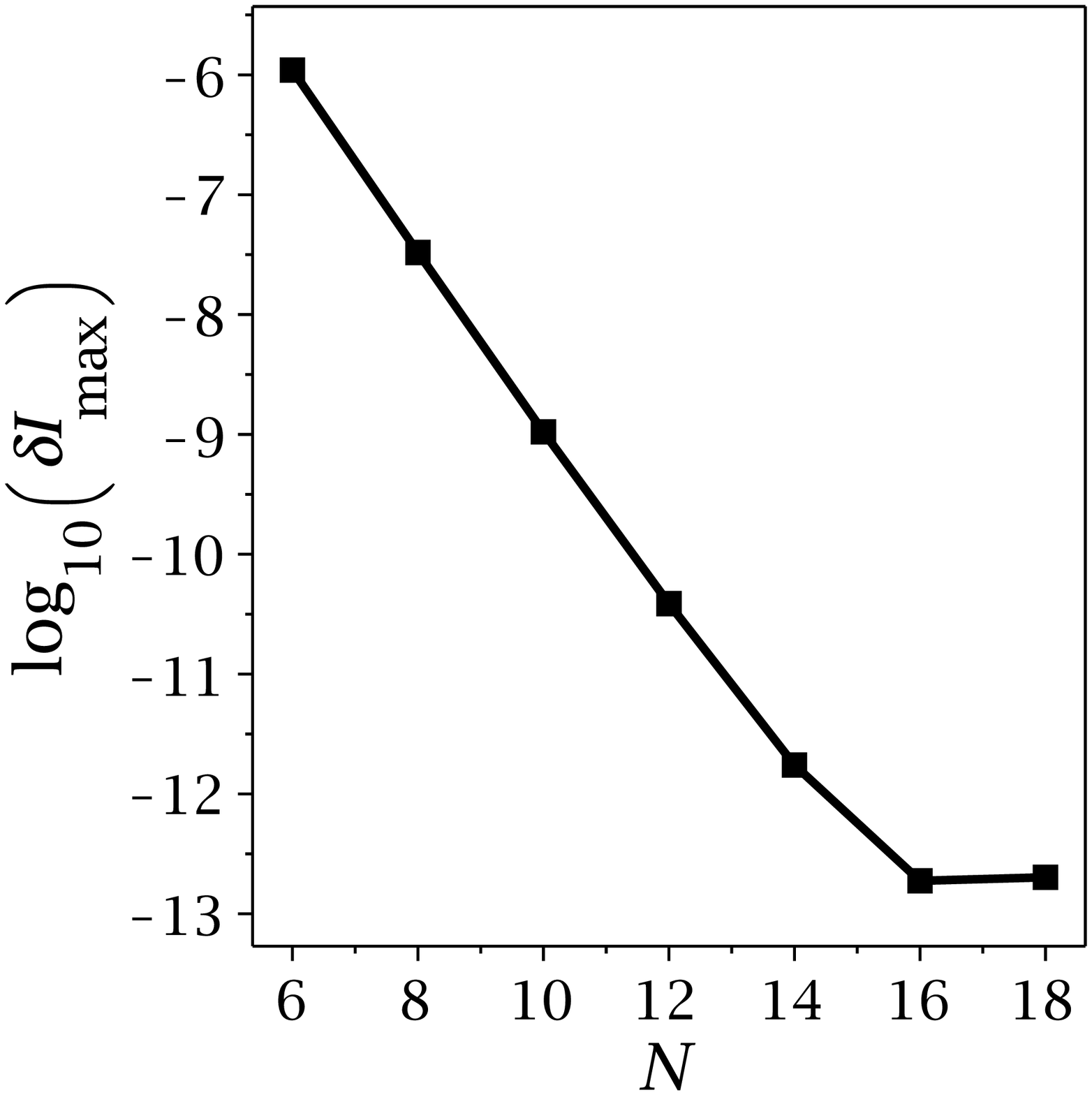}}
\rotatebox{0}{\includegraphics*[height=4.5cm,width=5.2cm]{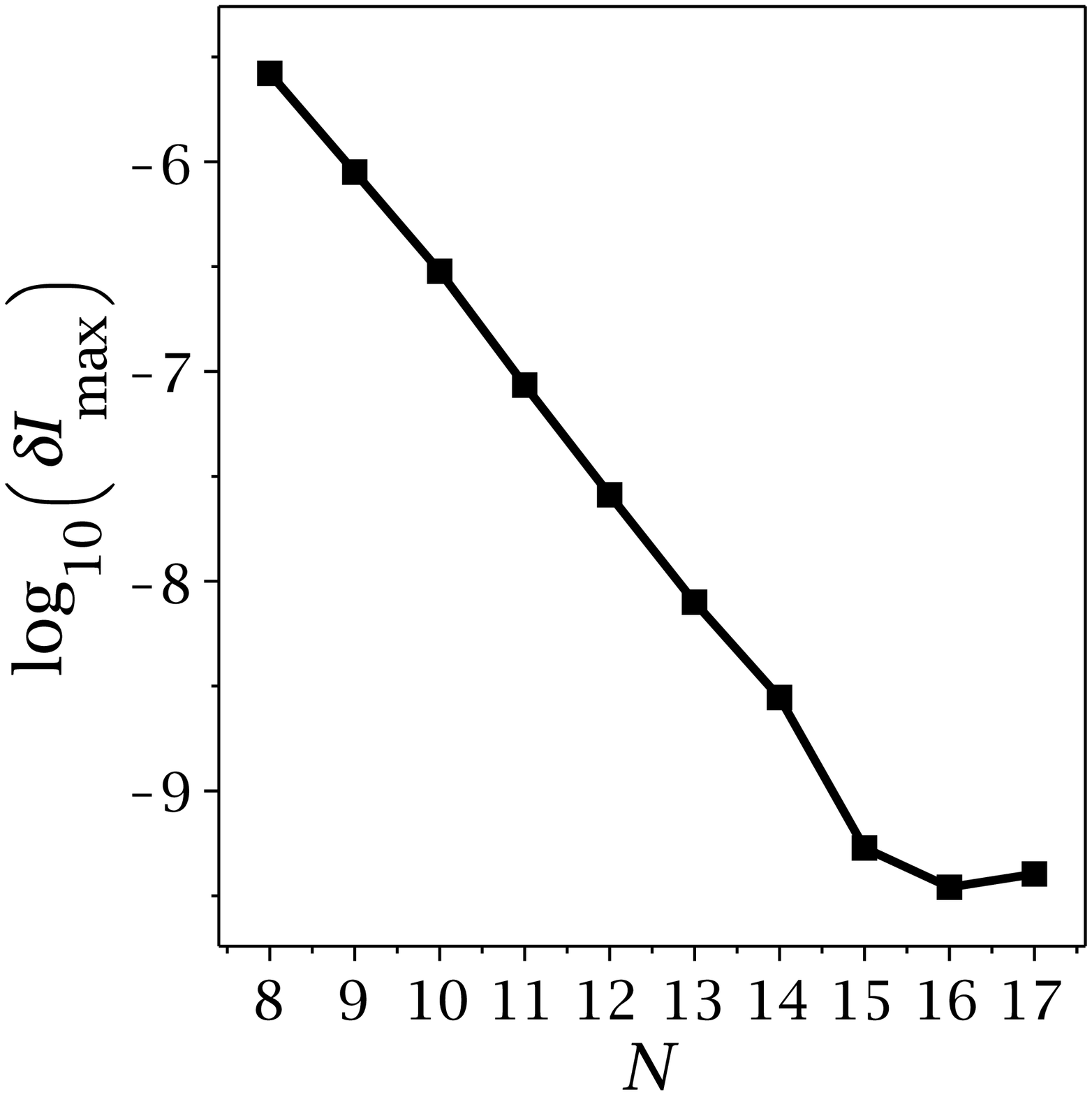}}
\caption{Plots of the maximum relative errors $\delta I$ after evolving the initial data (\ref{eq13}), (\ref{eq14}) and (\ref{eq15}) (from top to bottom). In all cases the exponential decay of the maximum error was observed as a result of the fast convergence of the code. The saturation due to the round-off error is shown for the two first cases and is achieved for different values of $N$.}
\end{figure}

In our numerical experiments we have determined the maximum deviations after evolving each family of  initial data for several values of $N$ as indicated in Fig. 1. In the first family of initial data~(\ref{eq13}) we set $e=0.6$ (note that a very small value for $e$ results in a weak perturbation of the stationary solution, which corresponds to $e=0$). For the second initial data~(\ref{eq14}) we have set $B_0=0.6,\zeta=0.1$ and used a Gaussian-like perturbation $f(x,\phi)=A_0 \sin^2\phi (1-x^2)^2\mathrm{e}^{-(x-0.3)^2}$, with $A_0=0.05$. The third family  of initial data (\ref{eq15}) is used with $P_0=1.0,\gamma=0.5$, while the perturbation is given by $g(x,\phi)=A_0 (1+\sin \phi) (1-x^2)^2\mathrm{e}^{-(x-0.1)^2}$, with $A_0=0.1$.  In Fig.~1 the log-linear plots of the maximum relative error versus the truncation order $N$ are presented for all three cases. Increasing $N$ produces an exponential decrease of $\delta I$ until the saturation due to the round-off error is achieved.

\begin{figure}[htb]
{\includegraphics*[scale=0.27]{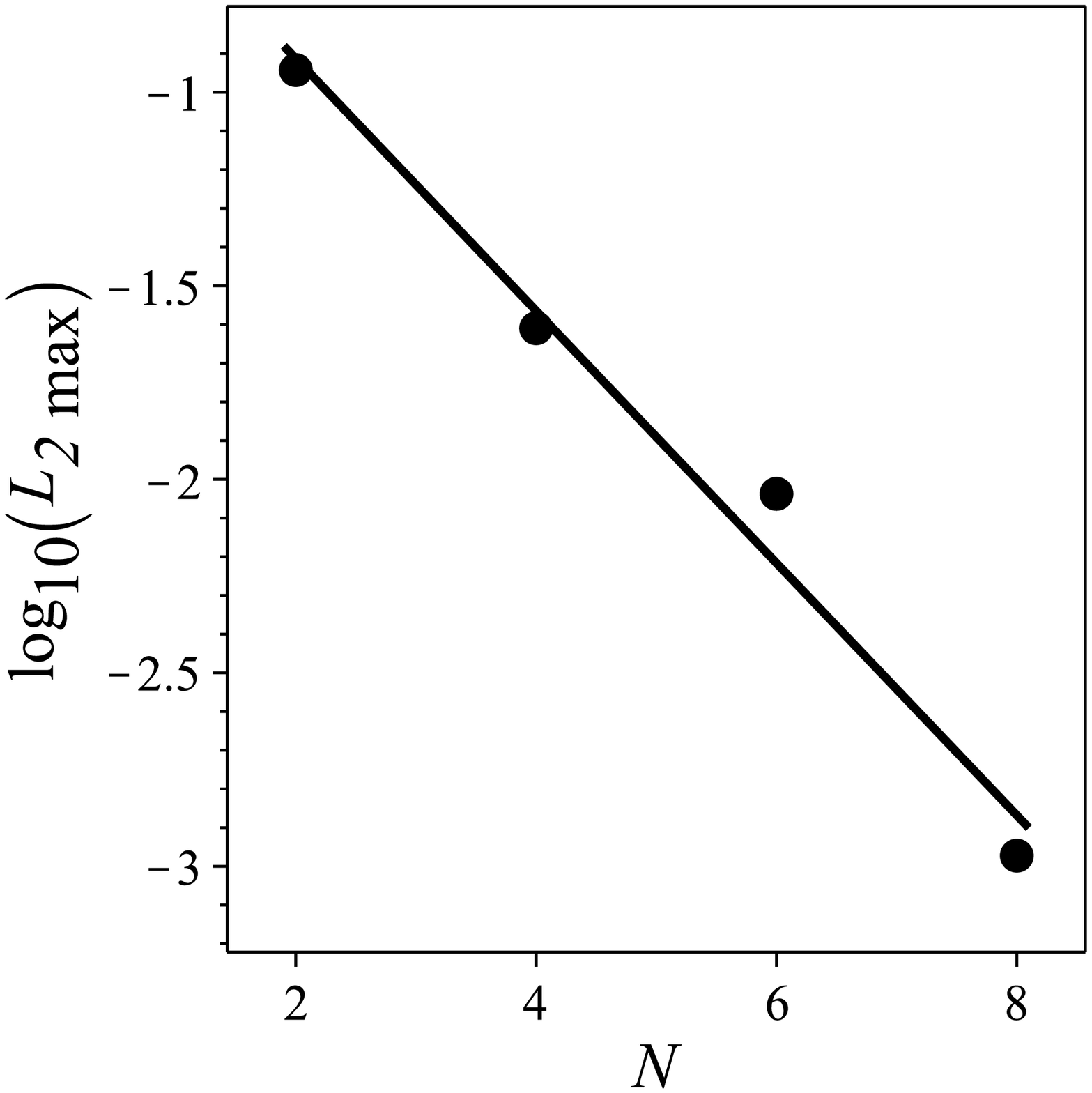}}
{\includegraphics*[scale=0.27]{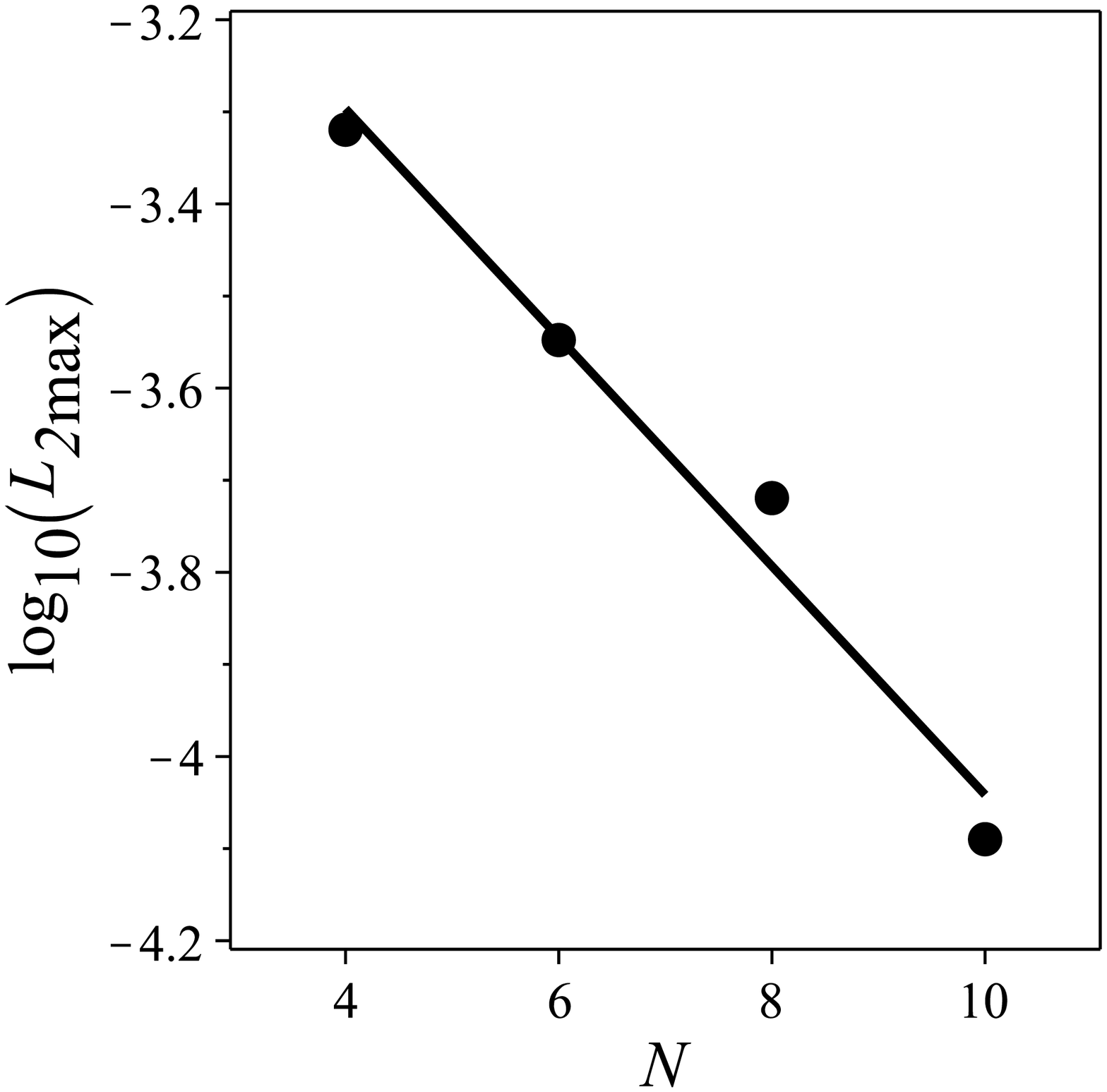}}
{\includegraphics*[scale=0.27]{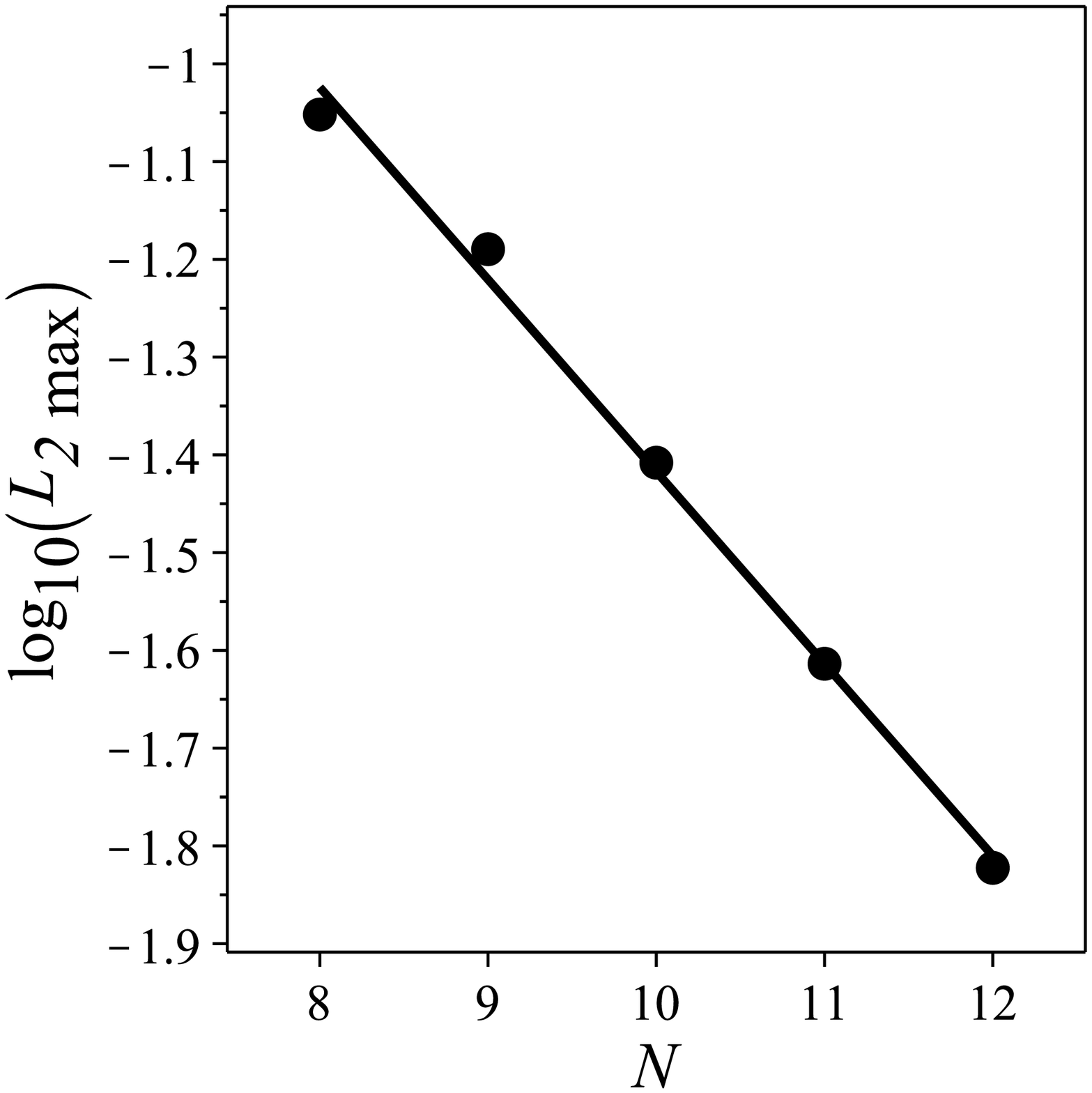}}
\caption{Plots of the maximum values of the $L_2$ norm (see Eq. (\ref{eq19})) with respect to the truncation orders  $N$ for the initial data (\ref{eq13}), (\ref{eq14}) and (\ref{eq15}) (from top to bottom). Again, spectral convergence is obtained.}
\end{figure}

Another code test is to evaluate the $L_2$ norm associated with the residual evolution equation~(\ref{eq7}) given by,

\begin{equation}
L_2(u) = \left(\frac{1}{4\pi}\,\int_{-1}^1\,\int_0^{2\pi}\,\mathrm{Res}_P^2(u,x,\phi) d\phi d x\right)^{1/2}.
\label{eq19}
\end{equation}

\noindent We have followed the evolution of $L_2$ for each set of initial data and several truncation orders and selected the corresponding maximum values. The results are shown in Fig.~2 and show spectral convergence.

The third code test is to reproduce correctly the Schwarzschild solution as the asymptotic configuration of RT spacetimes. In the axisymmetric case the asymptotic configuration can be either a black hole at rest or moving with constant velocity along the symmetry axis with respect to an inertial observer. The boost is a consequence of the total energy-momentum conservation when the flux associated with the gravitational waves is nonzero. The  net momentum of the gravitational waves is transferred to the source producing a boosted black hole along the symmetry axis. In the general case we expect that the resulting black hole will be moving in an arbitrary direction.

The structure of the initial data plays a crucial role in determining whether or not the black hole acquires momentum. Let us evolve the field equations taking the first set of initial data (Eq. (\ref{eq13})) with several values of the parameter $e$. In all cases the $a_{00}$ mode tends to a constant value while all other modes approach zero. The final configuration is $P_{\mathrm{final}} = a_{00} = \mathrm{constant}$,  which is identified as the Schwarzschild solution with mass-energy~\cite{oliv1,rt_austr,gonnakramer}

\begin{equation}
E_{\mathrm{BH}} = \frac{m_0}{4 \pi}\,\int_0^{2 \pi}\int_{-1}^1\frac{d \phi d x}{P^3_{\mathrm{final}}} = \frac{m_0}{a_{00}^3}. \label{eq20}
\end{equation}

\noindent In fact, the absence of any boost ($\gamma=0$) is a consequence of the symmetric pattern of gravitational wave radiation produced by the initial configuration (see Fig. 3).

We turn now to the less symmetric second and third sets of initial data. After several numerical experiments we have obtained the following asymptotic expression for the metric function $P$,

\begin{eqnarray}
P_{\mathrm{final}}(x,\phi) &&= \alpha_0 + \alpha_1 x+ \alpha_2 \sqrt{1-x^2} \cos \phi + \nonumber \\
&& \alpha_3 \sqrt{1-x^2} \sin \phi \label{eq21}
\end{eqnarray}

\noindent where $\alpha_j$, $j=0,1,..,3$ are determined numerically. This expression generalizes the stationary solution~(\ref{eq16}), representing a boosted black hole in an arbitrary direction with respect to an asymptotic observer as shown by Cornish~\cite{cornish2}. The parameters $\alpha_j$ are not independent but satisfy the constraint $\alpha_0^2-\sum_{j=1}^3\,\alpha_j^2=P_0^2=\mathrm{constant}$ imposed by Eq. (\ref{eq2}). This allows the following alternative choice for these parameters: $\alpha_0=P_0\,\cosh \gamma$, $\alpha_1=P_0 a \sinh \gamma$, $\alpha_2=P_0 b \sinh \gamma$, $\alpha_3=P_0 c \sinh \gamma$, which yields

\begin{eqnarray}
P_{\mathrm{final}}(x,\phi)& &= P_0[\cosh \gamma+ a \sinh \gamma x+ \nonumber \\
& &\sinh \gamma \sqrt{1-x^2} (b \cos \phi + c \sin \phi)],
\label{eq22}
\end{eqnarray}

\noindent and the relation,

\begin{equation}
a^2+b^2+c^2=1,
\label{eq23}
\end{equation}

\noindent that must be satisfied by the numerical solution asymptotically. Notice that $b=c=0$ implies $a=\pm 1$ recovering the axially boosted black hole~(\ref{eq16}), where the signs $\pm$ indicate opposing directions of movement along the $z$-axis. In the general case, the black hole is moving with respect to an inertial observer in an arbitrary direction determined by $a,b,c$, which can be identified as the direction cosines~\cite{rod_saa}. In this case the total mass-energy of the black hole is given by, %A useful quantity is the total mass-energy of the black hole:

\begin{eqnarray}
E_{\mathrm{BH}} = \frac{m_0}{4 \pi}\,\int_0^{2 \pi}\int_{-1}^1\frac{d \phi d x}{P^3_{\mathrm{final}}(x,\phi)} = m_{\mathrm{rest}}\cosh \gamma,
\label{eq24}
\end{eqnarray}

\noindent where $m_{\mathrm{rest}}=m_0 P_0^{-3}$ is the rest mass of the boosted black hole. The velocity of the hole is thus given by $\tanh \gamma$ as in the axisymmetric case. We have checked the relation~(\ref{eq23}) with the asymptotic numerical solution obtained after evolving the third family of initial data. The results are shown in Table~I with the deviation of $(a^2+b^2+c^2)_{\mathrm{numer}}$ from the exact value versus the truncation order $N$. Accordingly, for the smallest truncation order, $N=6$, the deviation is about one part in $10^{-5}$ confirming the excellent accuracy of the code.

\begin{table}[!ht]
\centering 
\begin{center}
\begin{tabular}{|c|c|}
\hline
$N$&$|(a^2+b^2+c^2)_{\mathrm{numer}}-1|$\\
\hline
\hline 6&$7.30 \times 10^{-5}$\\
\hline 7&$2.52 \times 10^{-5}$\\
\hline 8&$2.58 \times 10^{-6}$\\
\hline 9&$5.89 \times 10^{-7}$\\
\hline
\end{tabular}
\end{center}
\caption{The deviation $|(a^2+b^2+c^2)_{\mathrm{numer}}-1|$ evaluated for increasing truncation orders.}
\end{table}

\section{Wave forms and Efficiency of gravitational wave extraction. Non-frontal collision of two black holes}

One of the most important aspects of RT spacetimes is the presence of gravitational waves~\cite{rt}. As discussed in Refs.~\cite{oliv1,oliv3,oliv4}, the characterization of gravitational waves in RT spacetimes is based on the Peeling theorem~\cite{peeling}, for which the Weyl tensor is expressed as $C_{ABCD} \sim N_{ABCD}/r$ at large $r$ characterizing the wave zone. The relevant components of the Weyl tensor with respect to a semi-null tetrad basis~\cite{oliv1} are $C_{0303}=-C_{0202}=-D(u,x,\phi)/r+\mathcal{O}(1/r^2)$, where

\begin{eqnarray}
D(u,x,\phi) = \frac{P^2}{2} \frac{\partial}{\partial u}\,\Big[(1-x^2) \frac{P_{,xx}}{P} - \nonumber \\
\frac{1}{(1-x^2)} \frac{P_{,\phi \phi}}{P}\Big].
\label{eq25}
\end{eqnarray}

\noindent This function gives the time and angular dependence of the gravitational wave amplitude within the wave zone.

The structure and evolution of the angular pattern of gravitational waves is directly connected to the initial data. In the next numerical experiments we show the evolution of these patterns for different initial data by displaying a sequence of three dimensional polar plots of $D(u,\theta,\phi)$. In Fig. 3 the first and second sequences correspond to the initial data (\ref{eq13}) and (\ref{eq15}), respectively, where $D(u,\theta,\phi)$ is evaluated at several times. The symmetry in the angular distribution of $D$ in the first sequence implies that no momentum from the gravitational waves is transferred to the source. A distinct situation occurs for the second sequence in which a non-symmetric pattern of $D$ indicates the existence of a net flux of momentum carried away by gravitational waves. In this case, due to the conservation of momentum-energy, the end state is a boosted black hole described  by Eq. (\ref{eq22}). Another example of a non-symmetric angular pattern is shown by a sequence of two-dimensional plots in Fig. 5. Note the presence of lobes pointing in the directions of maximum emission of gravitational radiation. In this case the initial data represents an oblique collision of two Schwarzschild black holes (cf. Eq. (\ref{eq33})) whose end state is, in general, a boosted black hole.

\begin{widetext}
\begin{center}
\begin{figure}[htb]
\includegraphics*[scale=0.25]{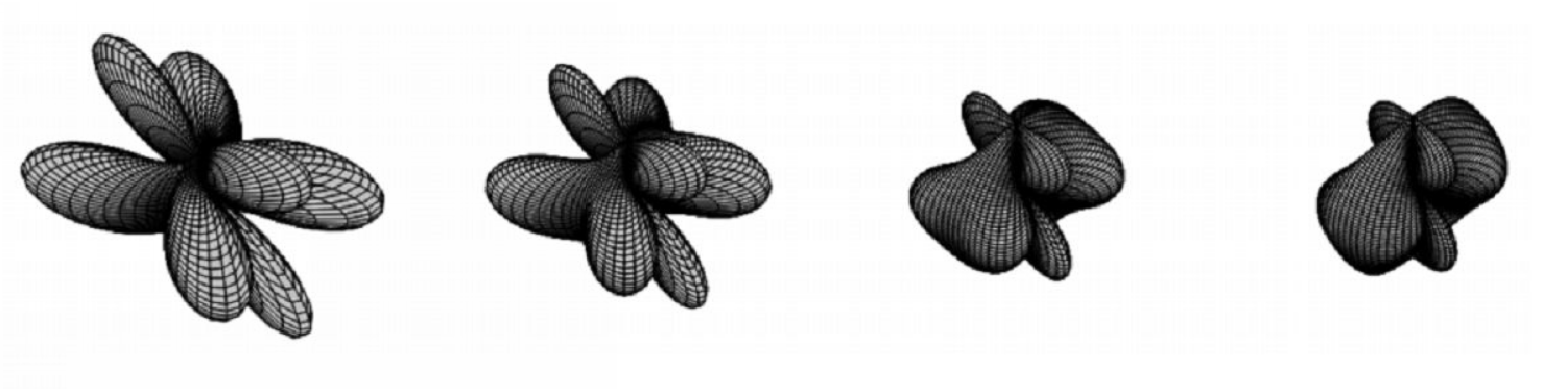}
\includegraphics*[scale=0.65]{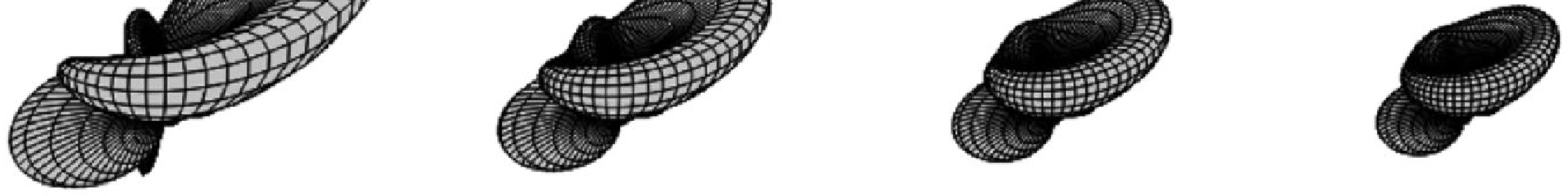}
\vspace{-16cm}
\caption{Sequence of three dimensional polar plots of $D(u,\theta,\phi)$ corresponding to the first set of initial data evaluated at $u=0,0.1,0.4,0.7$, and to the third set of initial data evaluated at $u=0.002,0.004,0.006,0.008$. The symmetric pattern in the first sequence suggests that no momentum is transferred to the remnant, while in the second sequence the lack of symmetry of the angular pattern indicates that the resulting black hole might be moving with respect to an asymptotic observer.}
\end{figure}
\end{center}
\end{widetext}
\vspace{-6cm}

\begin{figure}[htb]
{\includegraphics*[scale=0.35]{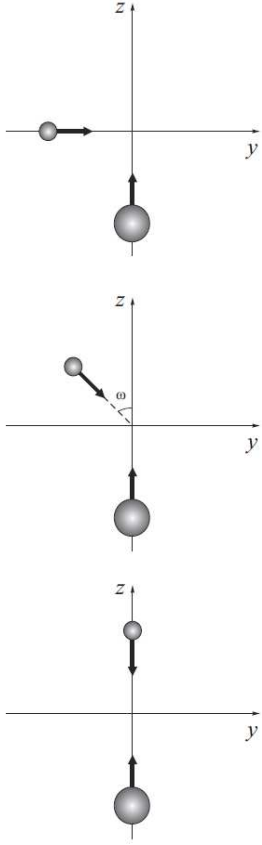}}
\caption{Illustration of the three situations representing (from top to bottom) orthogonal, oblique and head-on collisions of two Schwarzschild black holes.}
\end{figure}

\begin{figure}[htb]
{\includegraphics*[scale=0.3]{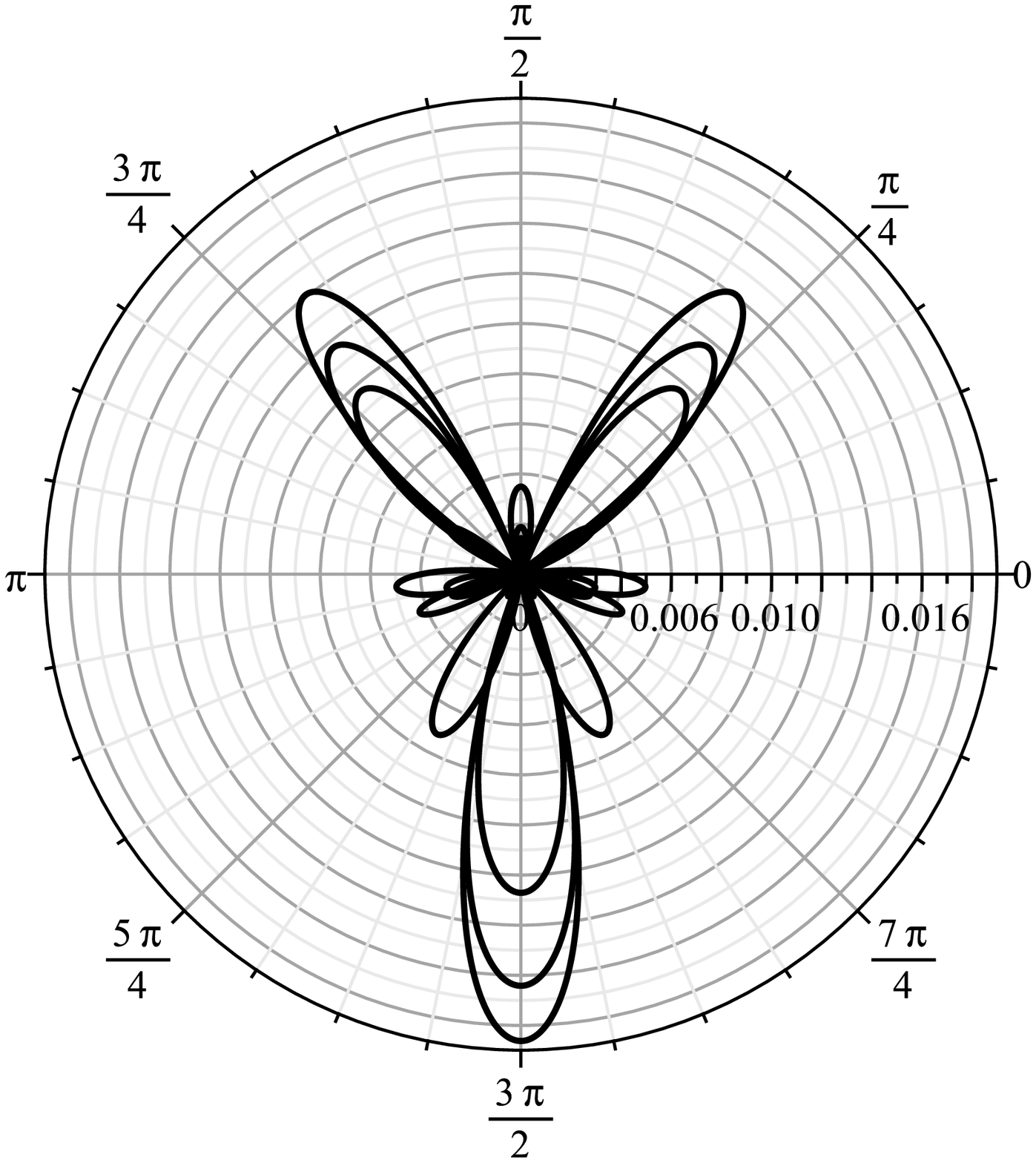}}
{\includegraphics*[scale=0.3]{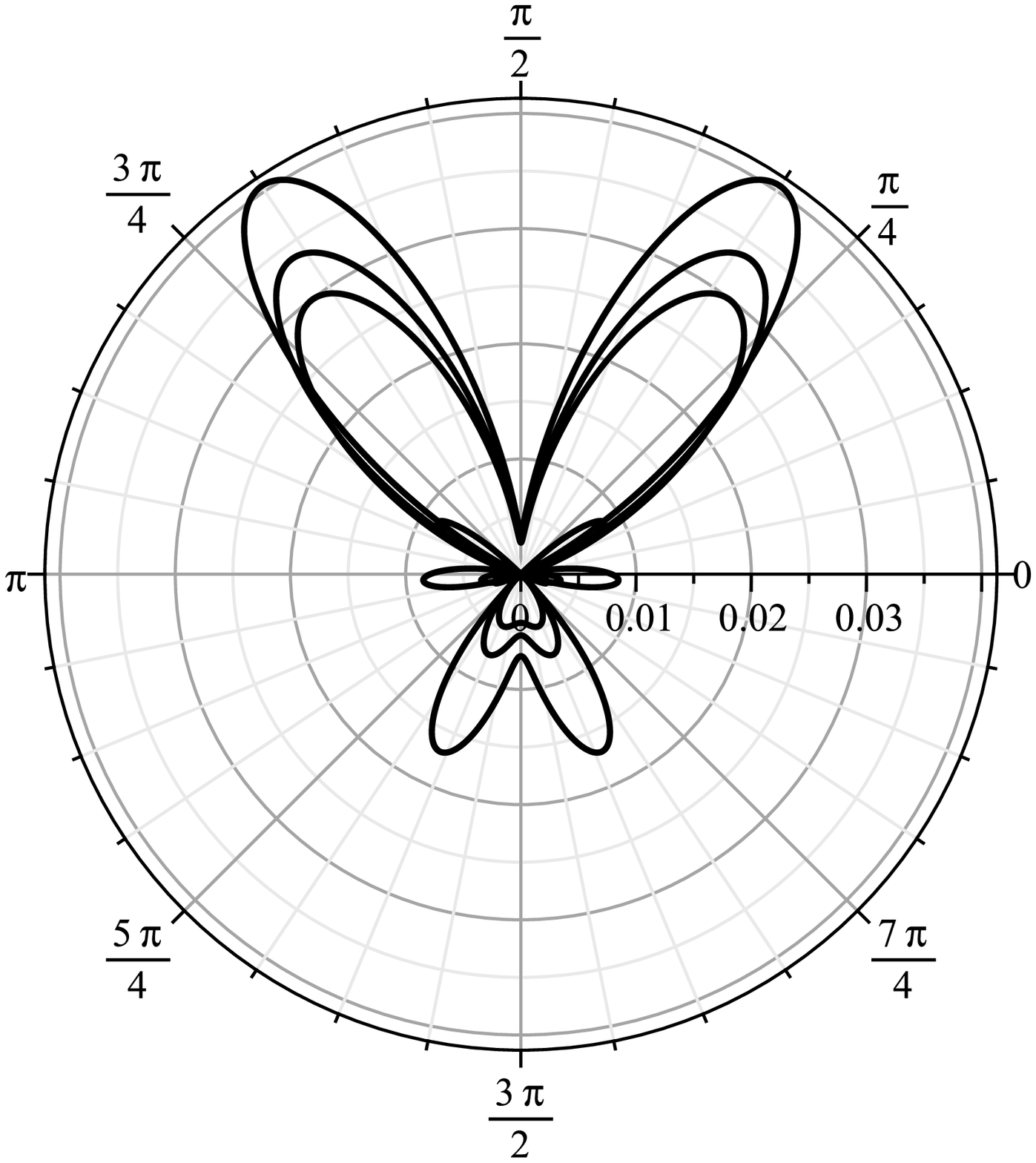}}
\caption{Illustration of the sequence of polar plots of $D$ projected on the planes $x=0$ ($u=0.02,0.10,0.26$) and $x=0.3$ ($u=0.02,0.10,0.18$) for the oblique collision. The angular pattern is dominated by lobes which indicate the directions of maximum magnitude of the gravitational wave emission. Note that the two dominant lobes open as the black hole is decelerated, forming a bremsstrahlung-like pattern.}
\end{figure}

The efficiency of gravitational wave emission measures the amount of mass-energy extracted during the evolution of RT spacetime with the eventual formation of a black hole. It is expressed as

\begin{equation}
\Delta=\frac{E_{\mathrm{I}}-E_{\mathrm{BH}}}{E_{\mathrm{I}}}=1-\frac{m_{\mathrm{rest}}\cosh \gamma_f}{E_{\mathrm{I}}}, \label{eq26}
\end{equation}

\noindent where $E_{\mathrm{I}}$ is the mass-energy of the initial system, $E_{\mathrm{BH}}=m_{\mathrm{rest}}\cosh \gamma_f$ is the energy of the remnant black hole and $\gamma_f$ is the final boost parameter. We can obtain analytical information about the efficiency of any process described by general Robinson-Trautmann spacetimes by analyzing the conservation of global four-momentum. Following Eardley~\cite{eardley_gw} we have,

\begin{eqnarray}
\label{eq27}
E_{\mathrm{I}}=E_{\mathrm{BH}} + E_{\mathrm{GW}} \\
\nonumber \\
p_{\mathrm{I}}=p_{\mathrm{BH}} + p_{\mathrm{GW}}, \label{eq28}
\end{eqnarray}

\noindent where $E_{\mathrm{GW}}$ and $p_{\mathrm{GW}}$ are the amount of energy and momentum carried by gravitational waves; $p_{\mathrm{I}}$ and  $p_{\mathrm{BH}}$ are the initial and black hole momentum magnitudes. These quantities are evaluated from the components of the four-momentum $\mathcal{P}^\alpha$ associated with RT spacetimes~\cite{rod_saa,moreshi},

\begin{equation}
\mathcal{P}^\alpha = \frac{m_0}{4 \pi}\,\int_0^{2\pi}\int_{-1}^1\frac{l^\alpha}{P^3(u,x,\phi)}\,dx d\phi, \label{eq29}
\end{equation}

\noindent where $l^\alpha = (1,\sqrt{1-x^2} \cos \phi,\sqrt{1-x^2} \sin \phi,x)$. Since the remnant is a black hole, it follows that $p_{\mathrm{BH}}=m_{\mathrm{rest}} \sinh \gamma_f$; also $E_{\mathrm{GW}}>p_{\mathrm{GW}}$ as a consequence of the radiated four-momentum being future pointing and timelike. This inequality is conveniently re-expressed using Eqs.~(\ref{eq27}) and~(\ref{eq28}) as,

\begin{equation}
\cosh \gamma_f - \sinh \gamma_f < \frac{E_{\mathrm{I}}-p_{\mathrm{I}}}{m_{\mathrm{rest}}} \equiv \beta \geq 0, \label{eq30}
\end{equation}

\noindent since to $E_{\mathrm{I}} \geq p_{\mathrm{I}}$. Now, from the initial data $P(u_0,x,\phi)$ $E_{\mathrm{I}},p_{\mathrm{I}}$ can be evaluated via Eq. (\ref{eq29}); also $m_{\mathrm{rest}}$ is determined from the conserved quantity $I$ \footnote{By calculating the conserved quantity $I$ from Eq. (\ref{eq17}) with $P$ given by Eq. (\ref{eq22}) we found $I=P_0^{-2}$. However, since $I$ has its value fixed by the initial data, it follows that $P_0=I^{-\frac{1}{2}}$ which determines the rest mass of the resulting black hole as $m_{\mathrm{rest}}=m_0/P_0^{-3}=m_0 I^{\frac{3}{2}}.$}. As a consequence, $\beta$ becomes determined after specifying the initial data. A direct manipulation of the above expression results in the determination of the lower bound of the boost parameter, or $\gamma_{\mathrm{f}}>\gamma_{\mathrm{min}}=\mathrm{arccosh} [(\beta^2+1)/2\beta]$. In this way, the lower bound of the remnant velocity $v_{\mathrm{f}} = \tanh \gamma_{\mathrm{f}}$, is given by,

\begin{equation}
v_f > v_{\mathrm{min}} = \frac{|\beta^2-1|}{\beta^2+1}. \label{eq31}
\end{equation}

\noindent From the lower bound of the final velocity it is possible to estimate the upper bound of the efficiency for any initial data,

\begin{equation}
\Delta_{\mathrm{max}} = 1- \frac{m_{\mathrm{rest}} \cosh \gamma_{\mathrm{min}}}{E_{\mathrm{I}}}. \label{eq32}
\end{equation}

As a brief application of these estimates, we study the efficiency of the non-frontal collision of two Schwarzschild black holes initially with the same speed $v=\tanh \gamma$ described by the following initial data,
\begin{eqnarray}
P(u_0,\theta,\phi) = \left(\frac{1}{\sqrt{P_+(\theta,\phi)}} + \frac{1}{\sqrt{P_-(\theta,\phi)}} \right)^{-2} \label{eq33}
\end{eqnarray}

\noindent where $P_{\pm}=P^{(0)}_{\pm}[\cosh \gamma + a_{\pm} \sinh \gamma \cos \theta +\sinh \gamma \sin \theta (b_{\pm} \cos \phi+c_{\pm} \sin \phi)]$ with $a_\pm^2+b_\pm^2+c_\pm^2=1$. Note that the particular case $b_\pm=c_\pm=0$, $a_\pm=\pm 1$ reduces to the initial data for two initially boosted Scharwzschild black holes with opposite velocities $\tanh \gamma$ along the symmetry axis \cite{oliv3}. In the general case the black holes have initially arbitrary directions determined by the parameters $a,b,c$. By conveniently setting these parameters we can study three situations illustrated by Fig. 4, that is, two non-frontal collisions and one frontal collision. In all situations the initial motion of the  first black hole is along the $z$-axis, with $a_+=1,b_+=c_+=0$, whereas the second black hole has in turn (i) $a_-=b_-=0,c_-=1$ corresponding to a direction orthogonal to the $z$-axis, (ii) $a_-=-\sqrt{2}/2,b_-=0,c_-=-\sqrt{2}/2$, and (iii) $a_-=-1,b_-=c_-=0$ for a head-on collision. In all numerical experiments the boost parameter was fixed as $\gamma=0.9$ which corresponds to the initial relativistic velocity $v \approx 0.716$, $P_+^{(0)}=1$ and $P_-^{(0)}$ is a free parameter that defines the ratio $\eta$ between the initial rest masses of the two black holes,

\begin{equation}
\eta = \frac{m^-_{\mathrm{rest}}}{m^+_{\mathrm{rest}}}=\left(P_-^{(0)}\right)^{-3}. \label{eq34}
\end{equation}

\noindent In table II we summarize the results for several values of $\eta$. Note that the frontal collision is the most efficient in terms of carrying away energy by gravitational waves. In line with numerical simulations based on more realistic scenarios, the efficiency of the extraction of energy via gravitational waves never exceeds a few percent of the initial energy. The results are also consistent with analytical estimates for both the lower bound of the final boost parameter and the maximum efficiency, as given by Eqs. (\ref{eq31}) and (\ref{eq32}) respectively. The values of the maximum boost parameter are displayed in Table II for each case.

\vspace{-1cm}
\begin{widetext}
\begin{center}
\begin{table}[ht]
\begin{center}
\begin{tabular}{|c|c|c|c|c|c|c|c|c|c|}
\hline
\multicolumn{1}{|c|}{}&\multicolumn{3}{c|}{Orthogonal}
&\multicolumn{3}{c|}{Oblique}&\multicolumn{3}{c|}{Head-on}\\ \hline
$\eta $&$v_{\mathrm{min}}$&$v_{f}$&$\Delta (\% )$&$v_{\mathrm{min}}$&$v_{f}$&$\Delta (\% )$&$v_{\mathrm{min}}$&$v_{f}$&$\Delta (\% )$\\
\hline $0.001$&$0.597$&$0.601$&$0.608$&$0.525$&$0.537$&$1.23$&$0.498$&$0.514$&$1.473$\\
\hline $0.008$&$0.557$&$0.564$&$0.911$&$0.437$&$0.456$&$1.97$&$0.383$&$0.411$&$2.428$\\
\hline $0.125$&$0.509$&$0.520$&$1.307$&$0.306$&$0.327$&$3.019$&$0.160$&$0.202$&$3.797$\\
\hline $0.296$&$0.501$&$0.511$&$1.384$&$0.269$&$0.294$&$3.233$&$0.077$&$0.121$&$4.084$\\
\hline
\end{tabular}
\end{center}
\caption{Efficiency and the final boost of the remnant black hole for several values of $\eta$.}
\label{table:abc}
\end{table}
\end{center}
\end{widetext}

\section{Discussion}

In this paper we have studied numerically the dynamics of general Robinson-Trautman geometries. We have constructed an efficient and accurate numerical code based on the Galerkin method that generalizes the code we have implemented for the axisymmetric case~\cite{oliv_code}. Spherical harmonics were used as the basis for the spectral expansions of the metric functions $P(u,\theta,\phi)$ and $\lambda(u,\theta,\phi)$ (cf. Eqs.~(\ref{eq4}) and~(\ref{eq5}), respectively). However, as can be seen from these expansions, an increase in the truncation orders $N$ and $M$ increases considerably the number of modes $a_{kj},b_{kj}$. The residual equations associated with the field equations were projected using the quadrature formulas~Eq.~(\ref{eq9}). The field equations~(\ref{eq2}) and~(\ref{eq3}) were reduced to a set of ordinary differential equations for the $(N+1)^2$ modes $a_{kj}$ together with equations for the modes $b_{kj}$, the values of $\lambda$, $P$ and their derivatives (see Eqs. (\ref{eq11}) and (\ref{eq12})) at the collocation points. This procedure allowed us to reach, without much computational effort, a truncation order of $N_{\mathrm{max}}=14$,  corresponding to 225 modes $a_{kj}$ and an equal number of independent modes $b_{kj}$. The code tests consisted of checking the conserved quantity I, monitoring the L2-norm associated with the residual
RT equation, and reproducing the correct asymptotic solution described by Eq. (\ref{eq21}) of a black hole moving with constant velocity in a arbitrary direction with respect to an asymptotic observer~\cite{cornish2}. These tests confirmed the excellent accuracy and exponential convergence of the code.

Another important result is the generalization of physically relevant initial data deduced for the axisymmetric case~\cite{oliv3,oliv4,oliv_rad}. Though we have not presented the details of these derivations here, we have established initial data representing the exterior gravitational field of a perturbed oblate (prolate) spheroid of matter, for which the perturbation can be a disruption inside the source. We have also generalized the initial data representing a perturbed axisymmetric boosted black hole. In both cases the asymptotic solution is a boosted black hole in an arbitrary direction determined by the parameters $a,b,c$ (cf. Eq. (\ref{eq23})).

We have presented two applications of the code. The first is a visualization of the angular pattern of the gravitational radiation defined in the wave zone (cf. Eq. (\ref{eq25})), whose structure depends on the initial data. The second concentrates on the efficiency of the conversion of the initial mass to gravitational waves. In this case we have set up the equations governing the conservation of global four-momentum taking into account that the remnant is a boosted black hole moving in an arbitrary direction with respect to an asymptotic observer. Analytical estimates of the minimum or lower bound of the boost parameter and the maximum efficiency were derived from the initial data.

Finally we point out that a thorough study of the non-frontal collision of two Schwarzschild black holes in the realm of RT spacetimes is a subject for further investigation. The initial data describing a non-frontal collision
of two Schwarzschild black holes with the same speed, given by Eq. (\ref{eq33}), generalizes the initial data established for the head-on collision presented in Ref. \cite{oliv3}. The numerical integration of the field equations (\ref{eq2}) and (\ref{eq3}) provides a useful theoretical probe for studying the consequences of the non-frontal collision of two black holes \cite{moreshi}. In our preliminary analysis we have confirmed that a head-on collision is the most efficient source for extracting gravitational radiation, although obviously being a rarer astrophysical event than a non-frontal collision. The evolution of the apparent horizon would also be worth investigating.

The authors acknowledge the financial support of the Brazilian agencies CNPq, CAPES and FAPERJ.

\end{document}